\documentstyle[12pt]{article}
\begin{document}

\author{G.K. Leontaris$^{(*)}$ \\
Centre de Physique Theorique, Ecole Polytechnique,\\
F-91128 Palaiseau, France\\
and\\
N.D. Tracas \\
Physics Department, National Technical University,\\
157 80 Zografou, Athens, Greece}
\title{Soft SUSY Masses and the Dynamical Determination of the Gravitino Mass}
\date{}
\maketitle

\begin{abstract}
We discuss in detail the possibility of determining dynamicaly the
gravitino mass $m_{3/2}$, which is related to the supersymmetry
breaking scale, within the minimal supersymmetric standard
model (MSSM). Using the complete MSSM spectrum, we minimize the vacuum
energy  including  one-loop corrections and a cosmological term of
${\cal O}(m_{3/2}^4)$ induced by the underlying fundamental theory.
We find that both terms are necessary to determine dynamically the
gravitino mass. Other useful constraints for the low energy
phenomenology are also obtained.
\end{abstract}

\thispagestyle{empty}
\vfill
\noindent IOA 313/95

\noindent NTUA 47/95

\noindent January 1995

\vspace{.5cm}
\hrule
\vspace{.3cm}
{\small
\noindent
$^{(*)}$On leave of absence from Physics Department, University of Ioannina,
451 10 Ioannina, Greece}

\newpage

It is widely believed that the only plausible solution to the gauge
hierarchy problem is N=1 local Supersymmetry
\cite{susy}.
The gauge hierarchy
problem arises from quadratically divergent one-loop corrections to the
effective potential, those being of the form $(\Lambda ^2Str{\cal M}
^2/(32\pi ^2))$, where $\Lambda $ is the momentum cut-off, while
\begin{equation}
\label{strace}Str{\cal M}^2(z,\bar z)=\sum_n(-1)^{2s_n}(2s_n+1)
				       m_n^2(z,\bar z)
\end{equation}
The sum is over all particles with
field-dependent masses squared $m_n^2$
and spin $s_n$. Since ${\cal M}^2$ contains also the
Higgs mass-squared, this term induces a divergent contribution
destabilising the hierarchy $M_W<<M_{Pl}$, where $M_{Pl}$ is the
Planck mass.

In the spontaneously broken N=1 local Supersymmetry the $Str{\cal M}^2$,
which appears as a coefficient of the one-loop quadratically divergent
contributions, is given in terms of the
field dependent gravitino mass $m_{3/2}$ by the formula
\cite{FKZ}
\begin{equation}
\label{strace2}Str{\cal M}^2=2Q(z,\bar z)m_{3/2}^2(z,\bar z)
\end{equation}
where the dimensionless function $Q(z,\bar z)$ depends on the fields $z$
and  $\bar z$ through the Ri\c{c}i tensor of the K\"ahler
manifold and the function $f_{ab}(z,\bar z)$ which determines the kinetic
terms of the vector supermultiplets as well as the gauge coupling
constants.

In the fundamental theory of quantum gravity the non--vanishing of
$Q(z,\bar z)$ would imply corrections to the effective potential of the
order ${\cal O} (m_{3/2}^2M_{Pl}^2)$ which cannot be cancelled by any
contribution of low energy physics. The gravitino mass is given by
\begin{equation}
\label{m32}m_{3/2}^2(z,\bar z)=\mid {\cal W}(z)\mid ^2e^{k(z,\bar z)}
\end{equation}
where ${\cal W}(z)$ is the superpotential. The value of $m_{3/2}$ is related
to the scale of supersymmetry breaking which should not be much larger than
the electroweak breaking scale.\ Since $m_{3/2}^2(z,\bar z)$ is field
dependent, its vacuum expectation value (vev) should arise from the
minimization of the potential. Then, quadratically divergent loop
corrections proportional to $Str{\cal M}^2 $ will induce either
$m_{3/2}\rightarrow 0$ (unbroken supersymmetry) or $m_{3/2}\rightarrow
M_{Pl}$, therefore destabilizing again the hierarchy.

A possible solution to the hierarchy problem requires the vanishing of
$Q(z,\bar z)$ which motivated the no-scale supergravity
models\cite{noscale}. A further step towards this problem has been taken
the last few years by going beyond the N=1 local Supersymmetry, the
Superstring theory. In the context of the latter, and in particular of
their four-dimensional version
\cite{4dstring},
the effective supergravity
theory is strongly restricted. It has been shown
\cite{FKZ}
that there
exist examples in supergravity theories preserving the general features of
the superstring underlying theory which predicts a vanishing ${\cal
O}(m_{3/2}^2M_{Pl}^2)$ contribution. Such theories, however, will still
leave a non vanishing contribution to the vacuum of the order ${\cal
O}(m_{3/2}^4)$ , which can be interpreted as a contribution to the
cosmological constant
\begin{equation} \label{dvcosm}\Delta V_{COSM=}\eta
(Q)m_{3/2}^4
\end{equation}
The energy scale dependent coefficient $\eta (Q)$ has a certain boundary
condition
on the unification scale. Its value there is dictated by the structure of
the `hidden' sector in the specific string model that has been chosen.

In reference
\cite{kpz},
the gravitino mass has been treated as a dynamical
variable. This would in turn imply that the low energy effective potential
should be minimized not only with respect to the vev's of the Higgs fields
but also with respect to $m_{3/2}$. It has been stressed that this term
cannot be absent in low energies as far as the gravitino mass is not taken
as an external parameter. On the contrary, its contribution is determined
by the evolution of the coefficient $\eta (Q)$ from the GUT scale down to
the low energies on the one hand, and the dynamical determination of
$m_{3/2}$ on the other hand.

In what follows we wish to analyse the above procedure in a realistic low
energy supersymmetric theory. We take as an example the Minimal
Supersymmetric Standard Model (MSSM) which is endowed with all the salient
features of an effective supergravity theory. We will  show that
under the very general characteristics of the above theories, the
$\eta (Q)$ is non-zero and negative at $Q\sim M_Z$, as long as
$m_{3/2}$ lies in the desirable range of 100 GeV to 1\ TeV. Moreover the
dynamical determination of the $m_{3/2}$ scale through the minimization
of the effective potential puts constraints of the scalar mass spectrum
of the theory.

We consider therefore the MSSM. Following the discussion above, the only
terms relevant to the potential (including quantum corrections) are the
following
\begin{equation}
\label{v1}V_1(Q)=V_0(Q) + \eta (Q)m_{3/2}^4+\frac 1{64\pi ^2}
Str{\cal M}^4(\ln  \frac{{\cal M}^2}{Q^2}-\frac 32)
\end{equation}
$V_0(Q)$ is the (R.G.E. improved) tree-level potential while the appearance
of the last term is due to the radiative corrections (at one-loop level)
and its inclusion is necessary in order to stabilize the minimization
procedure of the potential against $Q$
\cite{fabio}.

The evolution of the parameter $\eta (Q)$ is determined by a R.G.E. which
can be derived by demanding that the potential $V_1(Q)$ is scale
independent to the one-loop order, i.e.
\begin{equation}
\label{dv1/dt}\frac{dV_1(t)}{dt}=0\ ,\qquad t=\ln Q
\end{equation}
Since the above relation should hold for all values of the fields, in the
case where $v_1=v_2=0$, we have
\cite{kpz,klnz}
$V_0\mid _{v_i=0}=dV_0/dt\mid _{v_i=0}=0$,
thus
\begin{equation}
\label{deta/dt}m_{3/2}^4\frac{d\eta (t)}{dt}-\frac 2{64\pi ^2}Str{\cal M}%
^4\mid _{v_i=0}=0
\end{equation}
The above differential equation determines the value of $\eta (t)$ in terms
of $Str{\cal M}^4$ and the gravitino mass, once the initial values
of $\eta$ and of the mass parameters entering $Str{\cal M}^4$, at the
unification scale,
are known.

The initial value $\eta _G$, for example, is related in some specific models to
the
difference $n _B-n _F$ where $n _{B(F)}$ are the bosonic\
(fermionic) degrees of freedom after supersymmetry breaking.
An explicit derivation of the cosmological term, which can be identified
with the contribution $\eta (Q)m_{3/2}^4$ of (Eq.\ref{v1}), is given in
ref.\cite{Ignatios}.
In this treatment, the supersymmetry breaking scale is
related to the size of a large internal dimension $R$. It was found that
after the SUSY breaking, the one-loop contribution
to cosmological constant is of the order of $(\alpha _{String}/4\pi
R^4)(\eta _B-\eta _F)$. For $Z_{2}\times Z_2$ orbifolds, the gravitino
mass is $1/\surd 8R$ , thus for broken SUSY one estimate that $\eta_G\leq 0$.

The initial values of the scalar masses $\tilde m_i^2$,
the gaugino mass $m_{1/2}$ and of the $\mu$ parameter at the
unification scale $M_G$, can be
parametrized in terms of $ m_{3/2}$
\begin{equation}
\label{ksi's}\tilde m_i^2=\xi _im_{3/2}^2{\ ,\qquad }m_{1/2}^2=\xi
_{1/2}m_{3/2}^2 {\ ,\qquad }\mu_G = \xi_{\mu}m_{3/2}
\end{equation}
where the $\xi$--coefficients are of ${\cal O}(1)$ (calculable in
specific models).
Therefore,  the value of $\eta (Q)$ at any scale $Q<M_G$ is given by
\begin{equation}
\label{eta(Q)}\eta (Q)=\eta _G+\frac 1{32\pi ^2}\int^Q_{M_G}
Str \hat{\cal M}^4(Q',\xi _i,\xi _{1/2})\mid _{v_i=0}d\ln Q'
\end{equation}
where%
$$
\hat{\cal M}^4\mid _{v_i=0}=\sum_i(2s_i+1)(-1)^{2s_i}
\frac{m_i^4(Q)}{m_{3/2}^4} =\sum_i(2s_i+1)(-1)^{2s_i}
\left[ f_i(\xi_i,Q)\right] ^4
$$
and $f_i(\xi _i,Q)$ can be calculated from the RGE running of the masses.
Therefore, the parametrization of (Eq.\ref{ksi's}) renders the value of
$\eta (Q)$, obtained from (Eq.\ref{eta(Q)}), independent of $m_{3/2}$.

For a given set of $(\xi _\alpha,\eta _G)$, the $m_{3/2}$ value will be given
by the minimization condition of the low energy potential with respect
to $ m_{3/2}$ . This condition results to the equation
\cite{kpz}
\begin{equation}
\label{etamincond}V_1+\frac 1{128\pi ^2}Str{\cal M}^4=0
\end{equation}
The latter has been interpreted as defining an infrared fixed point of the
cosmological term, as it corresponds to the vanishing of the associated
$\beta $-function. It is a significant constraint that should be
satisfied  by the $m_{3/2}$ and $\xi _\alpha$ parameteres and the low energy
values of the gauge couplings involved in $V_0(v_1,v_2)$.

In order to exploit the constraint of (Eq.\ref{etamincond}), in the case
where the complete spectrum of the MSSM is taken into account, we need the
detailed $Q${--}dependence of all the relevant parameters. We start with
the classical tree-level potential which is given by
$$
V_0(Q)=(m_{H_1}^2+\mu ^2)\mid H_1\mid ^2+(m_{H_2}^2+\mu ^2)\mid H_2\mid
^2+m_3^2(H_1H_2+hc)
$$
\begin{equation}
\label{V0}+\frac{g^2}8(H_2^{\dagger }\vec \sigma H_2+H_1^{\dagger }\vec
\sigma H_1)^2+\frac{g^{\prime ^2}}8(\mid H_1\mid ^2-\mid H_2\mid ^2)^2
\end{equation}
The minimization of the $V_0$ potential with respect to $v_{1,2}$
leads to the well known conditions
$$
m_3^2=-\frac 12(m_{H_1}^2+m_{H_2}^2+2\mu ^2)\sin 2\beta
$$
\begin{equation}
\label{mincond}\frac 12M_Z^2=\frac{m_{H_1}^2-m_{H_2}^2\tan ^2\beta }
{\tan ^2\beta -1}-\mu ^2
\end{equation}
The conditions of (Eq.\ref{mincond}) allow us to write the tree level
potential in a simple\break
form, exhibiting its dependence on the $M_Z$ mass.
Substituting (Eq.\ref{mincond}) into (Eq.\ref{V0}) we get
\begin{equation}
\label{V0(vb)}
V_0(v,\beta )=-\frac 1{32}(g^2+g^{\prime 2})v^4\cos ^22\beta
=-\frac 1{8\pi }\frac{M_Z^4\cos ^22\beta }{(\alpha +
\alpha ^{\prime })}
\end{equation}
where $v=246$GeV. We can use the minimization condition (Eq.\ref{etamincond})
to determine
the required low energy value of $\eta (Q)$ as a function only of the
parameters  $\xi _{\alpha}$ and $\xi _Z=(M_Z/m_{3/2})^2$ and $\tan\beta$,
i.e. for $Q\sim M_Z$ we get
\begin{equation}
\label{etamz}\eta (M_Z)=\frac 1{8\pi }\left\{ \frac{\xi _Z^2\cos ^22\beta}
{\alpha +\alpha ^{\prime }}-\frac 1{8\pi}Str \hat{\cal  M}^4(\ln \hat{\cal
 M}^2-1)\right\}
\end{equation}
The above relation enables us to calculate the required low energy value
of  the cosmological coefficient $\eta (Q)$ for any set of the parameters
$\xi _\alpha$ choosing a phenomenologically  acceptable $m_{3/2}$ range. By
solving then the corresponding RGE for $\eta (Q)$, (Eq.\ref{deta/dt}),
we can determine a consistent range of values of $\eta $ at the
unification scale. Some general remarks concerning (Eq.\ref{etamz}) are
worth noting here.

First there is a positive contribution from the tree level potential which
depends on $\xi _Z$ and the angle $\beta $ . For very large $\tan \beta$,
this term becomes almost independent of $\beta $ as $\mid \cos 2\beta \mid
\rightarrow 1$.
As $m_{3/2}$ shifts to values much larger than $M_Z$, then
 $\xi_Z^2 \ll 1$ and the positive contibution becomes negligible.
There is a negative contribution, on the other hand, from
the supertrace dependence which finally leads $\eta (M_Z)$ to negative
values at $m_Z$.
Scalar mass and gaugino contibutions in the supertrace scaled by $m_{3/2}$
are independent of the latter, being functions only of the $\xi_\alpha$
parameters and the scale dependent gauge functions. Therefore, the main
$m_{3/2}$ dependence enters through the logarithmic terms of the form
$\ln({\tilde m}_i^2(Q)/Q^2)-1\equiv \ln{f(\xi_i,Q)}+\ln(m_{3/2}^2/Q^2)
-1$. Therefore, the $m_{3/2}$ value at which the minimum of the
potential occurs, is intimately related to these terms.

The calculation of $Str \hat{\cal  M}^4$ requires the
knowledge of the boundary conditions (b.c.) for the scalars at the GUT
scale, i.e. the knowledge of the $\xi _{\alpha}$ parameters. In the case
of universal b.c., for example, one has $\xi _i=\xi _0=m_0/m_{3/2}$ and
$\xi _{1/2}=m_{1/2}/m_{3/2}$, i.e. only two parameters in addition to
$\xi _Z$. However, in the general case of supergravity
theories $\xi _i$ are in general different (non-universality) and the
parameter space becomes more complicated. In addition, the RGEs for the
scalars should also contain the contribution of the U(1)-D terms which
plays a significant role for large deviations from the universality
condition  $\xi _i=\xi _0$.

The important fact of the above described approach is that, for a specific
supergravity or superstring model, up to an overall constant which can be
identified with the gravitino mass, all the $\xi _\alpha$'s are known. If in
addition the initial value of $\eta (Q)$ at $M_{G}$ is known, equations
(Eqs.\ref{deta/dt},\ref{etamincond}) can determine exactly the gravitino
mass.

In practice, it is not trivial to write down, at least for the moment, a
detailed spectrum of a realistic string model. Therefore, in the present
analysis we prefer to follow the above described procedure using the
general features of a supergravity theory. In this procedure, we treat
as free parameters the coefficients $\xi _\alpha$, varying them in a range
close to unity, and use the complete spectrum of the MSSM to predict a
consistent range of $\eta (Q)$ at the unification scale. This bottom-up
approach has as a prerequisite the knowledge of the gravitino mass
whose value is supposed to be determined dynamically. We know however
that, since supersymmetry breaking is related closely to the
$m_{3/2}$--scale, its value should be necessarily of the order of the
electroweak scale. Our purpose is then to show that under realistic
conditions and for a wide choice of the parameter space $\vec \xi
=\left( \xi_i,\xi _{1/2},\xi _\mu\right) $ there are some stable and well
defined predictions of the input value $\eta (M_{G})$ which can be
hopefully determined independently in specific string models. To put it
in another way, using all the possible information of low energy
physics, one can certainly support, or rule out, possible string
constructions.

In the present work we stick in the low $\tan \beta $ regime and prefer to
use semianalytic formulae to calculate the $Str$-contributions. To start
with, in the case of non-universal conditions at the GUT scale for the soft
terms, we generalize our previous
formulae\cite{leo} for the third generation of
Squarks which are the only one affected by the heavy top contribution.

As in ref \cite{kpz}, we prefer to restrict our analysis in the case of the
universal condition in the Higgs sector, although it seems interesting to
consider the more general case. However, working in the low $\tan \beta$
regime, the non universality in the Higgs sector, $m_{H_1}^0=m_{H_2}^0$,
is not expected to play a significant role, contrary to the case of large
$\tan\beta$ scenario. In the latter case, departure from
universality\cite{nonuni}  is sometimes necessary to avoid instabilities
in the low energy effective potential due to large negative corrections
to both Higgs mass parameters.
We give now the specific formulae which we are going to use.

The RGEs for the scalars receiving large $h_t$ Yukawa contribution are
\begin{equation}
\label{dmqdt}
\frac{d\tilde m_{Q_L}^2}{dt}=
\sum \frac{c_i^QM_i^2g_i^2}{8\pi ^2}-%
\frac{h_t^2}{8\pi ^2}\left(\tilde m_{Q_L}^2+
			 \tilde m_U^2+m_{H_2}^2+A_t\right) -
			 \frac 16\frac{\alpha _1}{2\pi }S
\end{equation}
\begin{equation}
\label{dmudt}
\frac{d\tilde m_U^2}{dt}=
\sum \frac{c_i^UM_i^2g_i^2}{8\pi ^2}-
\frac{2h_t^2}{8\pi ^2}\left(\tilde m_{Q_L}^2+
                          \tilde m_U^2+m_{H_2}^2+A_t\right) +
			  \frac 23\frac{\alpha _1}{2\pi }S
\end{equation}
\begin{equation}
\label{dmhdt}
\frac{dm_{H_2}^2}{dt}=
\sum \frac{c_i^HM_i^2g_i^2}{8\pi ^2}-%
\frac{3h_t^2}{8\pi ^2}\left(\tilde m_{Q_L}^2+
			  \tilde m_U^2+m_{H_2}^2+A_t\right) -
			  \frac 12\frac{\alpha _1}{2\pi }S
\end{equation}
where $S$, in the case of MSSM, is given by%
$$
S=m_{H_2}^2-m_{H_1}^2+\sum_{gen}(\tilde m_Q^2+\tilde m_D^2+\tilde m_E^2-
				 \tilde m_L^2-2\tilde m_U^2)
$$
The solution of the above system can be easily found through the solution
of the differential equation obeyed by the sum of the three masses
$u(t)=\sum \tilde m_i^2$, (where we have assigned
$\tilde m_1\rightarrow \tilde m_Q^2$,
$\tilde m_2\rightarrow \tilde m_{_U}^2$ and
$\tilde m_3\rightarrow m_{H_2}^2$),
\begin{equation}
\label{dudt}\frac{du(t)}{dt}=u_0(t)-\frac{6h_t^2}{8\pi ^2}u(t)
\end{equation}
where%
$$
u_0=\sum_j\sum_i\frac{c_i^jM_i^2g_i^2}{8\pi ^2}-\frac{6h_t^2}{8\pi ^2}
A_t\quad ,\qquad j=Q,U,H_2
$$
It is worth noticing here that the (Eq.\ref{dudt}) is independent of the
$S$ contribution, since the sum of the U(1) charges should be zero in the
term  $QUH_2$ ( invariance of the Yukawa Lagrangian under U(1$)$)$.$ Of
course, each individual mass gets a contribution from the $S$ term.
The solution of the differential equation is given by
\cite{leo}
$$
u(t)=\int_{t_0}^t u_0(t)dt-6\delta _A^2(t)-6\delta _m^2(t)
$$
Following closely the formalism of ref.\cite{leo} and
taking into account that the $A_t$ contributions are small, we can
write the solution of the (Eq.\ref{dmqdt}--\ref{dmhdt}) in the form
\begin{equation}
\tilde m_n^2=\xi _nm_{3/2}^2+C_n^i(t)\xi_{1/2}^im_{3/2}^2+C_n^S(t)S_0
m_{3/2}^2-n\delta _m^2(t)
\end{equation}
where the coefficients $C_n^i(t)$ are defined in ref.\cite{leo} and
$$
C_n^S=\left\{ -\frac 16,\frac 23,-\frac 12\right\} \frac
1{b_1}\left( \frac{\alpha _1(t)}{\alpha _1(t_G)}-1\right)
$$
$$
S_0 =\xi _3-\xi _{H_1}+\sum_{gen}(\xi _1+\xi _D+\xi _E-\xi _L-2\xi _U)
$$
where the differential equation obeyed by $S$, namely
$dS/dt=\alpha _1b_1S/(2\pi )$, has been used. In the following we will
stick in the
case $\xi _3=\xi _{H_1}$ (universality in the Higgs sector)  and that
all three $\xi_{1/2}$ are the same (universality in the gaugino sector).

The $Str$-term contibutions can be calculated now easily.
We should point out  that this calculation involves the $\mu $ parameter
of the superpotential, which is unknown at the $M_{G}$ scale. However,
in the bottom-up approach we are using here, the minimazation conditions
at $Q\sim M_Z$ determine  the value $\mu $ at this scale. Its value at
any scale can be obtained by generalizing (Eq.26) of ref.\cite{leo}
for the case for non-universal b.c., and evolve it using the relevant
renormalisation group equation.

A final issue we should discuss before we present our numerical
results, is the scale at which the required parameters should be
calculated. Indeed, as we shall show soon,
$\eta (Q)$
varies substantialy as the scale approaches $M_Z$ and its value in very
sensitive to the chosen scale.  For a gravitino mass close to the value
$M_Z$ it seems sensible to calculate all the relevant parameters at $Q\sim
M_Z$. If we seek however  solutions for $m_{3/2}\gg M_Z$, it would be
 appropriate to calculate  the relevant quantities at a scale
close to this value of $m_{3/2}$. Then, according to our program we define
as low energy value of $\eta (Q)$ that one obtained from the minimization
condition at $Q = m_{3/2}$ and  calculate the required initial condition
$\eta_G$ at $M_G$.

We start our numerical investigations with the renormalisation group of
the coefficient $\eta (Q)$. Using Eq.(\ref{eta(Q)}), in Fig.1a we plot the
coefficient $\eta (Q)$ using as initial value $\eta _G=0$, for three
characteristic  choices of the coefficients $\xi _{\alpha}$,
\break
$\alpha=i,\frac 12,\mu$.
By varying them in a reasonable range, we find that a crusial role
is played by the choice of the coefficient $\xi_{1/2}$. For each
particlular choice of $\xi_i$'s, we choose the value of $\xi_{\mu}$ so
as to ensure radiative breaking of the $SU(2)\times U(1)$
symmetry at the low energy scale.  From the three curves shown in
Fig.1a, the upper one corresponds to the choice $\xi_{1/2}=
\frac {1}{4}$, the middle to the case to $\xi_{1/2}= 1.8$, while  the lower
 to the value $\xi_{1/2}= 5$. We observe that the bigger the coefficient
$\xi_{1/2}$, the lower the value of $\eta (Q_Z)$ obtained for the same
initial condition $\eta_G$. This is of cource expected since larger
contibutions in the $Str{\cal M}^4$, result also to a bigger value of
$\eta (Q)$ through (Eq.\ref{eta(Q)}). It is clear from
(Eq.\ref{eta(Q)}) that a different initial condition $\eta_G$ will
result to a parallel shift of the  obtained curves by the same amount.
In Fig.1b we examine the sensitivity of the $\eta (Q)$ with respect to
the $\xi_i$ parameters for given $\xi_{1/2}=1.8$. We present three
cases where the parameter $S_0$ takes the values $3.6, 3.3, -1.6$.
Although we observe a significant variation of the $\eta (Q_Z)$ value
for the above choices, this is smaller than the one obtained by varying
$\xi_{1/2}$.  On the other hand, there is no obvious interrelation between
$\eta (Q)$ and $S_0$ values. The final $\eta (M_Z)$'s depent solely
on the specific choice of $\xi_i$'s. On the contrary, we find a rather
interesting correlation between $\eta (Q)$ curves and
the top Yukawa coupling.
 In Fig.1c  we plot curves for $h_{t_G}=1.8,2.6$ and $3.0$, while
fixing all $\xi_i$'s with $\xi_{1/2}=1.8$. As  can be read from the
the curves, the higher the top coupling the lower the $\eta (M_Z)$
value.

In Figs.2--4 we present our results from the minimization procedure
with respect to $m_{3/2}$, varying the value of $\eta (Q\sim M_Z)$ to
a range close to the one obtained by the minimization condition of
(Eq.\ref{etamincond}). In our calculations we use $M_G \approx 1.3\times
10^{16}$GeV,  $\alpha_G\approx 1/24.6$ and a SUSY scale close to
$m_{top}$. The  obtained top mass is $m_{top}\approx 175$GeV while we
take $\tan\beta\approx 1.8$.

In Fig.2 we plot the low energy effective potential
$V_1(M_Z) ${\it vs} $m_{3/2}$ for a selected case where
\[
\xi_{1/2}=.25,\quad\quad \xi_Q=\xi_U=2.5\quad \mbox{\rm and}\quad
          S_0=-3.5
\]
and three choices of $\eta_Z=-4,-2,-1$. The electroweak breaking
occurs at $Q\sim 450$GeV.
We notice in the graph that in the specific case
mentioned above, for $\eta_Z$ in the range $(-1,-3)$, the minimum of
$m_{3/2}$ is in the range $(150,550)$GeV.
Of course, such a low $\xi_{1/2}$ will result low masses for the
gauginos, in particular for the larger $\eta_Z$ values of the above range
which give the lower $m_{3/2}$ minimum.
In Table I we give the masses of the SUSY particles scaled with the
$m_{3/2}$ mass.

\newpage
\begin{center}
{\bf Table I}

\vspace{.5cm}
\begin{tabular}{|ccc|ccc|cc|cc|}
\hline
$M_1$&$M_2$&$M_3$&$\tilde m_Q$&$\tilde m_U$&$\tilde m_D$&
		  $\tilde m_{t_L}$&$\tilde m_{t_R}$&
		  $\tilde m_L$&$\tilde m_E$\\
\hline
0.21&0.41&1.36&1.75&1.52&1.58&1.69&1.18&1.54&0.85\\
\hline
\end{tabular}

\vspace{.5cm}
{\it The masses of the three gauginos and the other SUSY particles, scaled
with the $m_{3/2}$ mass, for the choices $\xi_{1/2}=.25$,
$\xi_Q=\xi_U=2.5$ and $S_0=-3.5$}
\end{center}

\vspace{.8cm}
In Figs.3a--b  we present the case where $\xi_{1/2}=1.8$ while
all other $\xi$'s are as before, for two
different scales namely $Q \approx M_Z$ and $Q \approx 250$GeV.
The parameter $\eta$ takes the values $(-20,-30,-40,-50)$.
Table II shows the obtained supersymmetric spectrum scaled again with
the $m_{3/2}$. The scale of electroweak breaking is
$Q\sim 280$GeV.  Since
now $\xi_{1/2}$ is higher we expect the SUSY masses to be heavier than
before.

\begin{center}
{\bf Table II}

\vspace{.5cm}
\begin{tabular}{|ccc|ccc|cc|cc|}
\hline
$M_1$&$M_2$&$M_3$&$\tilde m_Q$&$\tilde m_U$&$\tilde m_D$&
		  $\tilde m_{t_L}$&$\tilde m_{t_R}$&
		  $\tilde m_L$&$\tilde m_E$\\
\hline
0.56&1.11&3.65&3.55&3.35&3.37&3.32&2.73&2.33&0.98\\
\hline
\end{tabular}

\vspace{.5cm}
{\it The same as in Table I, for the choices $\xi_{1/2}=1.8$,
$\xi_Q=\xi_U=2.5$ and $S_0=-3.5$}
\end{center}

\vspace{.8cm}
In Fig.3a, $V(Q,m_{3/2})$ develops a minimum for
$n(M_Z)\approx (-30,- 50)$ with a corresponding range of
$m_{3/2}\approx (120-550)$GeV. In Fig.3b, the minimum is
obtained for larger $\eta (Q)$ values being now in the range
$\eta(M_Z)\approx  (-25,-35)$.
The
minimum at $m_{3/2} = 300$GeV corresponds to $\eta (M_Z)\approx -40$ in
the first case (Fig.3a) and to $\eta(250$GeV$)\approx -31$ for
the second (Fig.3b). Again
as $\eta$ shifts to lower values, the minimum of
$m_{3/2}\rightarrow \infty$.
Notice that within the above range, $ V_1(Q)$ is  stable
with respect to the scale $Q$ as expected.

Fig.4 represents a case with relatively large value of
$\xi_{1/2}=5.0$
and $\xi_{Q}=\xi_{D}=0.8$ and $S_0=-0.3$.
All the relevant parameters are calculated at  $Q=M_Z$,
while the curves correspond to $\eta =(-200,-250,-300,-350)$.
Finally we wish to point out that the cosmological coefficient receives
naturally small values close to zero only in the first case, namely
when $m_{1/2}\le m_{3/2}$. From this point of view, a vanishing
cosmological constant at $Q\sim M_Z$ would require a considerabe
fine tuning of the various parameters.

The three cases chosen above are in  correspondence with the
curves obtained from the renormalisation group running of the
$\eta$--coefficient.  Comparing the results with Fig.1,
it can be seen that  large positive $\eta_G\ge {\cal O}(100)$ values
are required  in order for the $\eta (Q)$ value obtained from the
RGE running to match  with the low energy $\eta$'s consistent with
the minimization condition.  The larger the value of $\xi_{1/2}$, the
higher the $n_G$ value required to obtain resonable $m_{3/2}$ values
dynamically.

In conclusion, we have discussed in detail the implications of the
minimization of the vacuum energy with respect to the gravitivo mass.
We have shown that the requirement of determining a
hierarchically consistent  gravitino mass dynamically, leads to useful
constraints in low  energy and Unification scale physics.
In particular,  we have seen that the existence of a $V$--minimum with
respect to $m_{3/2}$ necessitates the inclusion of the one loop corrections
and of the cosmological term $\eta (Q)m_{3/2}^4$, remnant from the
underlying supergravity or string theory.
Furthermore the minimization of the vacuum energy can naturally
lead to $m_{3/2}$ values at the order of the elecroweak scale
$m_{3/2}\sim (100-500)$GeV and acceptable supersymmetric mass spectrum, in
particular if $m_{1/2} > m_{3/2}$.
Further constraints are also put on the $\eta_Z$ parameter which can be
easily converted to constraints for the initial value of the
cosmological coefficient $\eta_G \equiv \eta (Q=M_G)$.
In particular, small $\eta_G$ values as required by specific
string models  are compatible with
$m_{1/2} \le m_{3/2}$ and small deviations from the universality
condition for the scalars. In this case a sparticle spectrum
compatible with the experimental bounds, requires
$m_{3/2}\ge (3-4)\times M_Z$.

It is interesting that the above minimization procedure may also
apply to other undetermined parameters of the standard model,
i.e.  Yukawa couplings and fermion masses \cite{kpz,kprz,bd}.


\vspace{.5cm}
We have benefited from discussions with I. Antoniadis, S. Dimopoulos
 C. Kounnas and F. Zwirner.
 G.K.L would  like to thank CERN Theory Division
for hospitality during the early stages of this work.
The work of N.D.T. is partially supported by a C.E.C. Science Program
SCI-CT91-0729.
\vfill
\newpage


\vfill
\newpage

\begin{center}
{\bf Figure Captions}
\end{center}

\vspace{1cm}
{\bf Fig.1} The running of the parameter $\eta(Q)$ with initial value
$\eta(M_G)=0$. In (a) we plot
$\eta$ for three different values of $\xi_{1/2}=\frac{1}{4},1.8$ and
5, with all other $\xi$'s fixed. In (b), keeping $\xi_{1/2}=1.8$ we
plot $\eta$ for three values of $S_0=3.6,3.3$ and $-1.6$. In (c), we
keep $\xi_{1/2}=1.8$, all other $\xi$'s fixed and we vary the initial
top Yukawa coupling $h_t(M_G)=1.8,2.6$ and 3.0.

\vspace{.5cm}
{\bf Fig.2} The potential $V_1(M_Z)$ as a function of $m_{3/2}$ for a
selected case where $\xi_{1/2}=1/4$, $\xi_Q=\xi_U=2.5$ and
$S_0=-3.5$. The three curves corespond to $\eta(M_Z)=-4,-2,-1$.

\vspace{.5cm}
{\bf Fig.3} As in Fig.2, with $\xi_{1/2}=1.8$. All other $\xi$'s are
the same. In (a) we plot the potential for $Q=M_Z$, while in (b) we
plot the potential for $Q=250$GeV.

\vspace{.5cm}
{\bf Fig.4} As in Fig.2, with $\xi_{1/2}=5$, $\xi_Q=\xi_U=0.8$ and
$S_0=0.3$. The curves correspond to $\eta=-200,-250,-300,-350$.


\begin{thebibliography}{99}
\bibitem{susy}
S.~Dimopoulos and H.~Georgi, Nucl. Phys. {\bf B193}(1981)150.
\\
H.~P. Nilles, Phys. Rep. {\bf 110}(1984)1;
\\
H.~E. Haber and G. ~L. Kane, Phys. Rep. {\bf 117}(1985)75;
\\
A. ~B. Lahanas and D. ~V. Nanopoulos, Phys. Rep. {\bf 145}(1987)1;
\\
S. Ferrara, ed., ``Supersymmetry'' (North-Holland, Amsterdam, 1987);
\\
F.~Zwirner, in ``Proceedings of the 1991 Summer School in High Energy
Physics and Cosmology'', Trieste, 17 June, 9 August 1991 (E.~Gava,
K.~Narain, S.~Randjbar-Daemi, E.~Sezgin and Q.~Shafi, eds.), Vol.~1,
p.~193.
\bibitem{FKZ}S.~Ferrara, C.~Kounnas, and F.~Zwirner,
Nucl. Phys. {\bf B429}(1994)589, and references therein.

\bibitem{noscale}
E.~Cremmer, S.~Ferrara, C.~Kounnas and D.V.~Nanopoulos, Phys. Lett.
{\bf B133}(1983)61.
\\
J.~Ellis, A.B.~Lahanas, D.V.~Nanopoulos and K.~Tamvakis, Phys. Lett.
{\bf B134}(1984)429;
\\
J.~Ellis, C.~Kounnas and D.V.~Nanopoulos, Nucl. Phys. {\bf B241}(1984)406
and {\bf B247}(1984)373.
\bibitem{4dstring}K.S.~Narain, Phys. Lett.{\bf B169}(1986)41;
\\
W.~Lerche, D.~L\"ust and A.N.~Schellekens, Nucl. Phys. {\bf
B287}(1987)477;
\\
H.~Kawai, D.C.~Lewellen and S.-H.H. Tye, Nucl. Phys. {\bf B288}(1987)1;
\\
I.~Antoniadis, C.~Bachas and C.~Kounnas, Nucl. Phys. {\bf B289}(1987)87.
\\
I.~Antoniadis and  C.~Bachas  Nucl. Phys. {\bf B298}(1988)1.
\bibitem{kpz}C.~Kounnas, I. Pavel and F.~Zwirner,
 Phys. Lett. {\bf B335}(1994)403.
\bibitem{fabio}G. Gamberini, G. Ridolfi and F. Zwirner, Nucl. Phys.
{\bf B331}(1990)331; C. Kounnas, in Properties of SUSY particles,
p.496, World Scientific, Erice procedings 1994.
\bibitem{klnz}S. Kelley, J.L. Lopez, D.V. Nanopoulos and A. Zichichi,
CERN preprint CERN-TH.7433.
\bibitem{Ignatios}I. Antoniadis, Phys. Lett. {\bf B246}(1990)377.
\bibitem{leo}G.K. Leontaris, Phys. Lett. {\bf B317}(1993)569;
\\
G.K. Leontaris and N.D. Tracas,
{\it Low energy thresholds and the scalar
mass spectrum in minimal supersymmetry}, CERN preprint CERN-TH./94,
(to appear in Phys. Lett. {\bf B}.
\bibitem{nonuni}L. ~E. ~Ibanez and D. L\"ust, Nucl. Phys.
{\bf B382}(1992)305;
\\
A. Lleyda and C. ~Mu\~noz, Phys. Lett. {\bf B317}(1993)82;
\\
T. Kobayashi, D. Suematsu and Y. Yamagishi, Kanazawa report, 94-06;
\\
D. ~Mataliotakis and H. ~P. Nilles, Munich preprint TUM-HEP-201/94;
\\
M. ~Carena and  C.E.M. ~Wagner, CERN preprint CERN-TH.7393/94;
\\
N. Polonsky and A. Pomarol, Pensylvania preprint UPR-0627T,1994;
\bibitem{kprz}C.~Kounnas, I. Pavel, G. Ridolfi and F.~Zwirner,
 CERN-TH. preprint, in preparation.
\bibitem{bd} P. Binetruy and E. Dudas, LPTHE Orsay 94/73.

\end{thebibliography}
\end{document}